# Diffraction of transmission light through triangular apertures in array of retro-reflective micro-prisms

Yizhou Tan[1], Han Chen[2]

*1. College of Mechatronic Engineering and Automation, NUDT*

*2 College of Opto-Electronic Science and Engineering, NUDT*

*National University of Defense Technology (NUDT)*

*Changhsa, Hunan, 410073, CHINA*

*\*Corresponding author:* tanyizhou@126.com

**Abstract**: The array of micro-prisms was described by model of multi-period blazed gratings consisting of triangular apertures. The origins of hexagram-shaped diffraction patterns were interpreted based on multiple-beam interference and diffraction array theorem. The relation between zonal /line ghost fringes and imperfectly fabricated array structures was analyzed. Geometrical performances (e.g., the dihedral angle of micro-prism) were tested by measuring the features of diffraction patterns of samples from three retro-reflective sheeting manufacturers.

## 1. Introduction

The application of micro-prisms arrays ranges from traffic light back-reflectors to the security printing industry (e.g., Kinegrams for banknotes) [1-3]. Micro-prisms array is also promising in liquid crystal display and chip-to chip /intra-chip optical interconnections.

Diffraction patterns generated by array of corner- cube (c.c.) prisms have been studied [2-6]. For example, patents [2, 3] utilize diffraction effects to spread the retro-reflected light over larger cone (> 0.5 °) by improved prismatic structures. Diffraction pattern were observed by array of corner cubes on the (100) crystal planes of silicon [5]. However, those studies focused on the diffraction of reflective light, and no analysis of the transmission light through array of c.c. prisms has been reported as far as the author knows.

In this work, the diffraction features were studied in the transmission manner of micro-prisms array. When a collimated beam traverses the retro-reflective sheeting, a hexagram-shaped diffraction pattern was observed. We solve an inverse problem — determination of unknown geometrical parameters of micro-prisms, such as dihedral angle and optical axis orientation of micro-prisms, by measuring the diffraction patterns of c.c. array.

On the prismatic surface of the retro-reflective sheeting, the groove angle is approximately 70.53°. Adequate metrology for such faceted surfaces with steep slope is lacking [5-9]. Our experimental results show that diffraction pattern analysis is an available tool for measurement of microstructure with high local slope surfaces.

The unusual diffraction phenomena, such as additional lines and zonal fringes in the regular diffraction patterns, have been studied experimentally. The results indicated that both additional lines and zonal fringes can provide more valuable information for revealing the geometric imperfections in a micro-prisms array.

## 2. The imperfectly fabricated micro-prisms

The retro-reflective sheeting typically comprises an optically clear layer with one smooth surface, and the opposite prismatic surface. More than 5000 micro-prisms are densely positioned within one squared centimeter. As shown in Fig.1a, the micro-prism is a tetrahedron consisting of three mutually perpendicular faces. And three dihedral angles are equal to each other, i.e. ∠ADO=∠BEO=∠CFO. The dihedral angle errors may be caused by the apex point offset or displacement of base face.

In the reflection mode, the incident direction of a ray is denoted by unit vector $-\vec{Q}$, and its final direction is denoted by $\vec{r}$. For a perfect c.c. reflector, vector $\vec{r}$ is parallel to $-\vec{Q}$. For the imperfect one, there must be a deviation of emerging beam $\vec{T}$. According to [9,15], if the dihedral angle α, β, γ deviate from the theoretical value (~54.33°), and a, b, c are the normals to the three mirror's faces, the direction of the emerging beam could be calculated by Equation (1),

$$\vec{T} = \vec{Q} + 2\vec{Q} \times (+\alpha \cdot a - \beta \cdot b + \gamma \cdot c) \qquad (1)$$

Two types of imperfect structures are shown by Fig 1b, 1c respectively: In the first case, if the three dihedral angle errors have the same scale δ, Equation (1) can be simplified into an empirical formula (2). The divergence Δ of refracted beam is related to the dihedral-angle offset δ by

$$\Delta \approx 4\sqrt{2/3} \cdot n\delta \qquad (2)$$

As an example, if the three dihedral angle errors δ≈1" and index n≈1.6, the direction deviation of retro-reflected beam $\vec{T}$ is approximately 5.3".

In the second case, the base triangle (dotted Δ ABC) is an isosceles triangle, but not a regular triangle. The normal of base face is deviated from the plumb line. As shown in the Fig 1c, the tilted tetrahedron prism is equivalent to the combination of perfect c.c. prism with an optical wedge.

The Eqs (1) and (2) indicated that the dihedral angle errors made bad influences on the retro-reflective performance of micro-prism. The literatures [7-9] measured the dihedral angle of prism based on the interference principle. However the Twyman-Green interferometer is not suitable for vibration environment. In our work we adopted a new approach based on diffraction patterns to measure the dihedral angle (see Fig. 2 and Appendix). The experiment results show that the diffraction patterns of micro-prisms are not sensitive to vibrations in workshop.

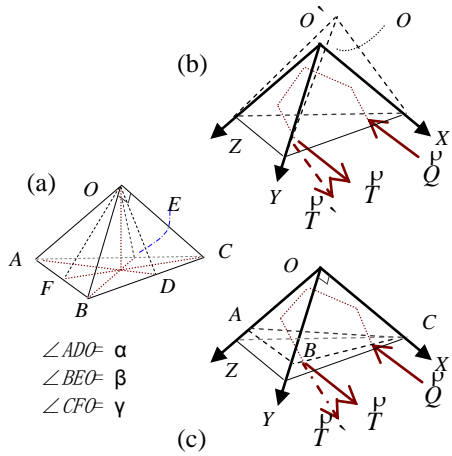

Fig.1 The retroreflective prism
(a) Perfect pyramid-shaped prism; (b) and (c) were imperfectly fabricated

## 3. Features of diffraction patterns

The schematic diagram of experimental setup is shown in Fig. 2. Retro-reflective sheeting is illuminated from its back surface (non-retro-reflecting side) by collimated laser ($\lambda \approx 633$nm, 5mW, ~3 mm in diameter). Fraunhofer patterns are observed by lens in far-field, and Fesnel patterns are observed in near-field.

Some new experimental phenomena are observed. Their features are different from the regular diffraction patterns reported by previous literatures. It is noteworthy that: (1) The hexagram-shaped image (Fig.2 b) is composed of six bright spots and nine diagonal lines, but patterns in [5,6] does not appear any diagonal line except bright diffraction spots; (2) There are multiple sets of distinct interference fringes modulated by hexagram diffraction envelope, as shown in Fig. 3d, which are alike the patterns diffracted by triangular apertures in [12,13].

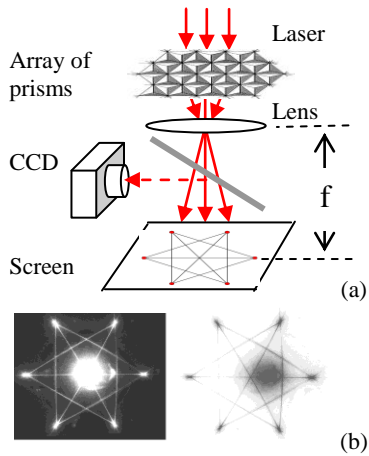

Fig. 2 Diffraction experiment in the transmission model
(a) Micro-prisms array under collimated laser illumination, (b) Photographs of diffraction patterns

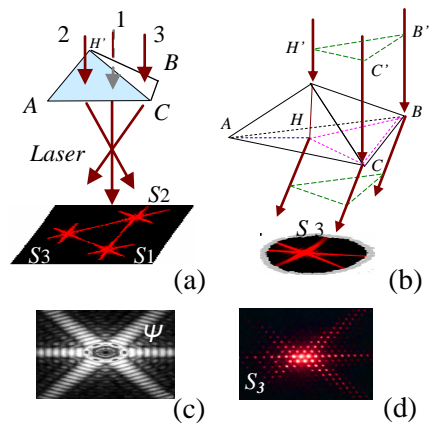

Fig. 3 Refraction and diffraction in a prism
(a) The parallel light was separated into three refraction beams, (b) Laser traveling through aperture $\triangle BHC$;
(c) Diffraction pattern formed by aperture $\triangle BHC$ (theoretical image $S_3$), (d) Experimental photo $S_3$.

The refraction and diffraction process in a prism is shown in Fig.3. Firstly the incident parallel light beam was refracted in three different directions (see Appendix). After traveling through the triangular pyramid, it was separated into three sub-beams and formed three diffraction patterns $S_1$, $S_2$, $S_3$ on the screen.

In transmission manner, c.c. array is a multi-period microstructure consisting of triangular apertures. Furthermore, it is a six-direction blazed prismatic grating with hexagonal array layout (when index≈1.6, blazed angle $\theta_4$≈31°, see Appendix).

(1) **Diffraction features of triangular aperture** Equation (3) is a statement of the array theorem, which says that the total aperture function $A(x, y)$ may be formed by convolving the individual aperture function with an appropriate array of Delta functions $A_\delta$. Hence

$$A(x, y) = A_i(x', y') \otimes A_\delta \quad (3)$$

Where $A_i(x', y')$ is the individual aperture function applicable to each transmission aperture, and the i*th* aperture has coordinates ($x_i + x', y_i + y'$) in the (*x, y*)-system. In this work, typical aperture function $A_i(x', y')$ is a obtuse triangle, and three obtuse triangle apertures in single prism are essentially identical (isosceles △*BHC*, △*CHA* and △*AHB* in Fig.3)

Distribution of diffraction energy was calculated numerically based on Huygens' Principle for single obtuse triangle aperture [12, 13]. The experimental pattern (Fig.3d) is similar in profile to theoretical patterns (Fig. 3c), but the experimental pattern contains multi-beam interference fringes originated from multi-beam interference of periodical apertures in c.c. array.

(2) **Six-pointed star patterns generated by a pair of prisms** In Fig.3a the collimated laser beam, traveling through a prism, divided into three refraction beams (denoted as $1^\#$, $2^\#$ and $3^\#$). Each beam generated a diffraction spot denoted as $S_1$, $S_2$, $S_3$ respectively. One prism generated a diffraction sub-pattern shaped in regular triangle △$S_1S_2S_3$. And the △$S_1S_2S_3$ contained three bright lines, which were the unique characteristics for triangle aperture's diffraction.

As Fig. 4c shown, bi-prism is the basic elements of an array. There are four kinds of bi-prisms. Each bi-prism contains adjacent prisms at opposite orientation. One bi-prism will certainly generate a diffraction image shaped like a six-pointed star, which can be regarded as a overlapping of dual sub-patterns ($\triangle S_1S_2S_3$ and $\triangle S_4S_5S_6$ in Fig. 4a, 4b).

The total six-pointed star pattern on the screen was the result of coherent superposition by large number, $N$, diffraction sub-patterns of bi-prisms in the array. The distribution function of interference apertures is Delta functions $A_\delta$ in Eq. (3).

(3) **The origins of diagonal lines** As shown in Fig.5, a set of adjacent apertures with opposite orientation ($O_{ai}$ and $O_{bi}$,) is responsible for generation of connecting line $S_{ai}$-$S_{bi}$ in diffraction pattern. Bright main diagonal line $M_1'N_1'$ on the screen is the result of constructive interference of multi apertures located on the one-dimensional periodical structures $M_1N_1$. A hexagon prismatic array includes three types of one-dimensional gratings, which produce three main diagonal lines, oriented respectively 0°, 120° and 240° on the circumference.

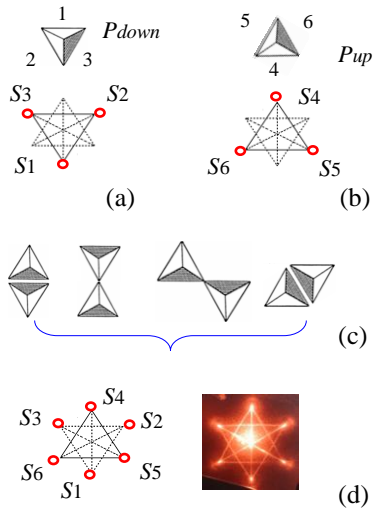

Fig. 4 Relationship between transmission apertures and its diffraction patterns
(a) and (b) Triangular sub-pattern diffracted by three apertures of single prism; (c) Four possible kinds of combination of adjacent prisms in c.c. array; (d) Six-pointed star pattern diffracted by six apertures of bi-prism

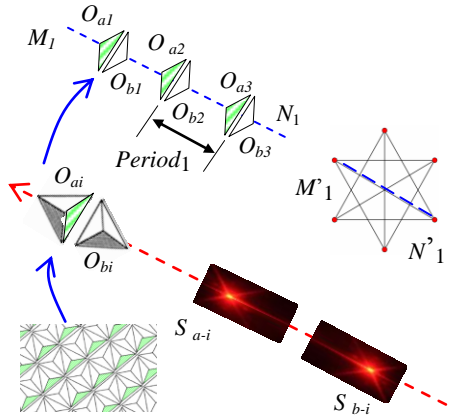

Fig. 5 Diagonal line formed by a 1-D prismatic grating
($O_{ai}$ and $O_{bi}$: adjacent apertures have opposite orientation, $M_1N_1$: one-dimensional structure constituted by periodical apertures, $M'_1N'_1$: diagonal line in diffraction pattern)

## 4. Experiments

### 4.1 Geometrical parameter of microstructure

Commercially available samples (from three manufacturers) are measured under the same testing conditions, and their hexagram-shaped diffraction patterns are shown in Fig.6. The dual orientation retro-reflective sheeting is a mixture of two groups of prisms, and each group has respective optical axis orientation [16-18]. Angles of cube axis orientation in Table 1 are measured for sample $1^{\#}$--$3^{\#}$ by the method of ray tracking [see Appendix, 14, 15].

(1) For a perfect c.c. prisms ($P_1$ and $P_2$), the V-shaped groove angle $\varepsilon_1 \approx 70.53°$. For sample $1^{\#}$, the experimental value $\varepsilon_1 \approx 70.63°$, and its diffraction pattern is a regular hexagram (Fig. 6c), which reveals that this sample is composed of single-orientation micro-prisms.

(2) Both sample $2^{\#}$ and $3^{\#}$ generated deformed hexagram patterns, which revealed that the orientation of prism axis in one group was not parallel to that of other group. The included angle between prism axes of two groups in c.c. array is proportional to relative displacement of geometric center of two diffraction sub-patterns (triangles $P_3$, $P_4$ and $P_5$, $P_6$ in Fig. 5d, 5e).

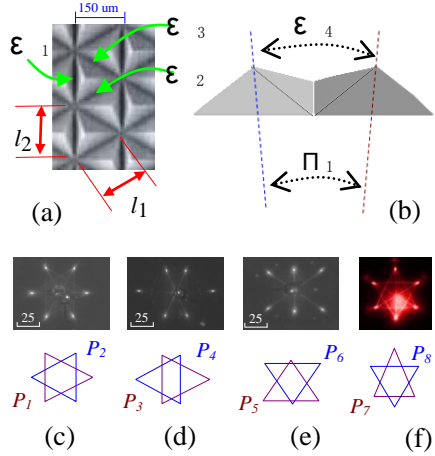

Fig. 6 Diffraction patterns of retro-reflective sheeting

(a) Photograph of corner-cube array, (b) Schema of dual-orientation micro-prism structure, (c)---(e) Diffraction patterns related to the variations of axis orientation of cube prism (samp1#, 2#, 3#, wavelength 0.63um). (f) Deformed diffraction pattern related to the dihedral angle of sample 4# ($P_7$: isosceles triangle, $P_8$: equilateral triangle)

Table 1 Geometrical parameters of retro-reflective sheeting

| Sample *No.* | 1# | 2# | 4# |
|---|---|---|---|
| Material Index | 1.541 | 1.466 | 1.574 |
| Refractive cosine $\theta_1$ $\cos(\theta_1)$ | 0.579 | 0.637 | 0.522 |
| Refractive cosine $\theta_1$ $\cos(\theta_2)$ | 0.576 | 0.433 | 0.674 |
| Groove angle $\varepsilon_1, \varepsilon_2$ ($\varepsilon_1=\varepsilon_2$) | 70.63° | 62.75° | 80.44° |
| Groove angle $\varepsilon_3$ | 70.63° | 89.71° | 58.38° |
| Ratio I of side $k'(L_2/L_1)$ | 0.990 | 1.144 | 0.847 |
| Ratio II of side $k(L_2/L_1)$ | 0.991 | 1.170 | 0.864 |
| Relative error $\Delta(k'-k/k')$ | 0.1% | 2.2% | 2.0% |

\* Symbols $\theta$, $\gamma$, $L_i$ are defined in Fig. 6, \*\*$L$:150~270um
  Laser wavelength 0.63um

The normal value of dihedral angle β (internal angle) is approximately ~54°20' for a perfect retro reflective prism (Fig.1, 2, 6c) [16-18]. There is a superposition of two sub-patterns (isosceles and equilateral triangle in Fig. 6f, sample $4^{\#}$). The diffraction sub-pattern shaped in isosceles triangle is related with the deviation of the dihedral angle $β_1$, which is caused by fabrication error. Dihedral angle $β_1 \approx 54°40'$ (experimental value), while other two dihedral angles $β_2$, $β_3$ for a same micro prism keep approximately 54°44'.

## *4.2 Ghost diffraction image*

The geometrical deviation of micro-prism array is one of the quality indicators. Manufacturers usually test this indicator to evaluate the performance of a retro-reflective sheet. Both Fig. 7b and Fig. 4d are diffraction patterns of high-quality retro-reflective sheeting. Their features are: (1) Its diffraction envelope is symmetrical; (2) it contains distinct multi-beam interference fringes (generated by array apertures with regular periodicity).

We introduce "Ghost" concept to explain the unusual diffraction phenomena, such as deformed hexagram-shaped diffraction patterns, by using earlier ghosts theory of one-dimensional grating for reference [10, 11]. In this paper, ghost fringes are defined as false, blur or deformed images in a diffraction patterns. As shown in Fig.7, 8, the Ghost fringes were mixed with regular hexagram patterns diffracted by c.c. array without geometrical imperfection. The measurement results for four types of ghost fringes were summarized in table 2.

Table 2. Imperfections in the c.c. array

| Number | 5# | 6# | 7# | 8# |
|---|---|---|---|---|
| Orientation Deviation | ±10' | +10' | < 10' | +40' |
| Ghost Lines/Zonal | 2 | 1 | 1~2 | Zonal |

In Fig. 7c one normal diagonal line and two additional lines were observed in horizontal direction. Those additional lines, i.e. the Ghost line in diffraction pattern, reveal that there were imperfectly fabricated micro-prisms in the sample $5^{\#}$. In detail, the normal apertures (isosceles triangle) were replaced in certain proportion by asymmetrical apertures owing to fabrication error. As Fig. 7b` shown, the isosceles triangle aperture generates symmetrical diffraction pattern. However, an obtuse triangle aperture generates deformed (or asymmetric) diffraction pattern as Fig. 7c`.

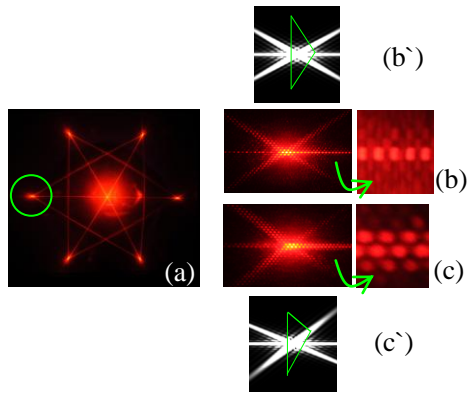

Fig. 7 Patterns diffracted by triangle apertures
(a) Diffraction pattern of c.c. array, (b) Generated by perfect apertures; (c) Pattern with ghost lines generated by imperfect apertures in sample $5^{\#}$ (wavelength 0.63um); Schematic diagram of pattern diffracted by isosceles triangle aperture (b`) and obtuse triangle aperture (c`)

There are one horizontal ghost line in Fig. 8c, and two ghost lines in Fig.8d. Those experimental phenomena provide a collateral evidence for evaluating the product quality: sample $6^{\#}$ is better than sample $7^{\#}$ in manufacture quality.

As a special case, there is only one horizontal diagonal line in Fig. 8b; it is still a ghost line. The false line reveal that the geometric periodicity in horizontal direction was good for all 1D-grating structures. However the prismtic axes of all 1D-dimensional gratings deviated synchronously to the normal direction owing to the fabrication error [13-15].

It is known that the triangle apertures in 1D-dimensional periodical structure are responsible for the generation of diagonal line in Fig 5. There are two additional horizontal lines associated with main diagonal line in Fig.8d (sample 7#). Those ghost lines reflect indirectly that orientation axes of some micro prisms were not parallel to adjacent 1D-dimensional prismtic gratings. In fact, it is the fabrication error that caused deviations of prismtic axes between two groups: one group keeps the normal orientation, and the other group is imperfectly fabricated with random error in its micro-prism axis.

The zonal ghost in Fig. 8e was generated by overlapping of multiple diffraction patterns with random difference. The zonal ghosts (together with visibility of interference fringes) reveal that retro-reflective sheeting (sample 8#) contains imperfect micro prisms in certain proportion.

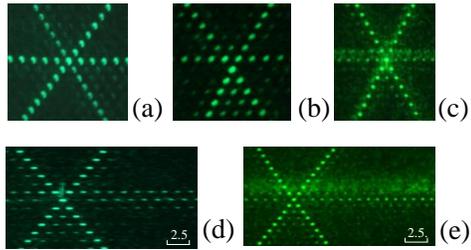

Fig. 8 Ghost fringes in diffraction patterns

(a) Diagonal lines of perfect 1D-dimensional periodical structures, (b) Single line is a ghost; (c), (d) Ghost lines associated with main diagonal line in sample 6#, 7#; (e) Zonal ghost in sample 8# (wavelength 0.5 3um)

## *4.3 Discussion*

Methods for testing micro-prisms array were in-depth studied by [7, 8]. The 25 and 50 μ m micro-corner cubes arrays were measured by a digital holographic microscope in transmission configuration with a 60x objective. An interferometric technique was reported in [8], which combined plane wave illumination with an index matching liquid to measure microstructures. Both [7] and [8] have the advantage to measure transparent high aspect-ratio of micro corner-cube and display 3-D mapping image of c.c. array. In principle, the above technique can measure micro-prisms one by one.

In our experiment the illumination area of laser beam is larger than the c.c. prism, e.g. laser diameter ≦ 3 mm, prism's scale ≅ 200um in Fig 2. Our testing approach can not supply 3-D mapping image for single micro-prism in an array. This can be considered as shortcomings by comparison with digital imaging technology in [7, 8].

In the other hands, it needs to be pointed out that the areas of retro-reflective sheeting are usual 15m×15m ~ 45.7m×1220mm. Both MZ interferometer and profilometer can not completely satisfy the request for testing those commercial products with a large area. Testing times become a bottleneck for holographic microscopy [7]. However, our testing method is suitable for testing samples in large area, because our diffraction instruments can measure more than hundred micro-prisms in each frame of CCD image.

The statistical testing principle in this work may after all be accepted concerning the higher efficiency for measurement in workshop environment. Firstly, the statistical information contained in diffraction pattern has potential application for on-line monitoring geometrical defects in micro-prisms array. Secondly, because the patterns diffracted by array apertures have the translation invariance, testing in vibration environment is in practice possible and feasible.

## 5. Summary

The diffraction testing method can provide statistical measurement data of least $10^2$ prisms in c.c. array in one action. Typical parameters (e.g., the dihedral angle, inclined angle of optical axis of micro prism etc) were measured by analyzing the features of hexagram-shaped diffraction patterns. The angle measuring accuracy $\leqq 0.5°$ for the groove angle in array. The relative errors of length $\leqq 2.2\%$ for micro-prisms in dimension 25-150um.

The "Ghosts" concept is helpful for evaluating the fabrication error, such as randomly distributed geometric imperfections in retro-reflective sheeting and periodicity degradation.

The above mentioned measurement results are related to the geometric index of c.c. array. The retro-reflectivity of retro-reflective sheeting is an optical index. In the next step, the problem to be solved is to set the relation between geometric and optical indexes by reference to official test and calibration standards for commercial products with fabrication defects.

## Appendix: Measuring the dihedral angles

Using the notation shown in Fig.9, let us trace the laser ray $L_i$ along the LQRGS path. Firstly consider the ray incident on the upper surface at angle $\theta_1$, and the ray traverses within the △ *ADO* section plane (see Fig. 1). Finally it emerges from the second surface at angle $\theta_4$. From the Snell's law of refraction, we can write

$$n_0 \cdot \sin\theta_1 = n_1 \cdot \sin\theta_2 \quad (A1)$$

$$n_1 \cdot \sin\theta_3 = n_0 \cdot \sin\theta_4 \quad (A2)$$

Where the index of air $n_0 = 1$ and $\theta_2 = \theta_1 + \theta_3$. Using the formulas of trigonometric relation, we solve the angle $\theta_1$, gives

$$n_0 \cdot \sin\theta_1 = n_1 \cdot (\sin\theta_1 \cdot \cos\theta_3 + \cos\theta_1 \cdot \sin\theta_3)$$

$$n_0 \cdot \tan\theta_1 = n_1 \cdot (\tan\theta_1 \cdot \cos\theta_3 + \cos\theta_1 \cdot \sin\theta_3)$$

In the experimental setup in Fig.2, 9, we keep the incident direction of laser $L_i$ perpendicular to the base plane, therefore $\theta_1 = \alpha_i$.

$$\alpha_i = \theta_1 = \tan^{-1}[n_1 \cdot \sin\theta_3 / (1 - n_1 \cdot \cos\theta_3)] \quad (A3)$$

Where $\theta_3 = \sin^{-1}[(\sin\theta_4)/n_1]$. The formula (A3) is the basic tools for calculating the dihedral angles and ghost lines by measuring the angle $\theta_4$ in this paper.

**Discussion 1**: the error transfer relationship between $\theta_1, \theta_4$ and $\alpha_i$

Because $\alpha_i = \theta_2 + \theta_3$, its differential results $d\theta_2 = -d\theta_3$. By the differential calculus of Equations (A1), (A2), we obtain

$$\sin\theta_1 \cdot d\theta_1 = n_1 \cdot cns\theta_2 \cdot d\theta_2 \quad \text{(A4)}$$

$$n_1 \cdot \cos\theta_3 \cdot d\theta_3 = cns\theta_4 \cdot d\theta_4 \quad \text{(A5)}$$

Division of one equation by the other gives

$$d\theta_4 / d\theta_1 = -(\cos\theta_1/\cos\theta_2) \cdot (\cos\theta_3/\cos\theta_4)$$

Typical parameters for our experiments are $n_1 \approx 1.6$, $\theta_1 = 45°$, $\theta_2 \approx 26.2°$, $\theta_3 \approx 18.8$, $\theta_4 \approx 31°$. In general, the transfer relationship of measurement errors is estimated by $d\theta_4 \geq (0.87)d\theta_1$. Because $\alpha_i$, $\theta_4$ errors and $\theta_1$ offset have the same magnitude, calibration of the incident angle $\theta_1$, i.e. keeping parallel between the $L_i$ direction and the plumb line, is the key to improve the measurement accuracy of the dihedral angle $\alpha_i$.

**Discussion 2**: the influence of the sample's thickness on the error of $\theta_4$

There is a pair of prisms $P_i$ and $P_j$ with symmetrical structure in Fig.9. The sample's thickness equals to the sum of base-layer's thickness $d_i$ and prism's height $h_i$. In the prism $P_i$, we measured the refracted angle $\theta_4$ with a CCD camera. In this far-field imaging case, both $d_i$ and $h_i$ have virtually no contribution to the error of $\theta_4$ value.

In near-field measurement case, the index $n_1$ base-layer's thickness $d_i$ is error sources, which should not be ignored. In prism $P_j$ we first measure the length $v_j$ and the horizontal deviation $k_j$, and then calculate the angle $\theta_4$ by the formula $\theta_4 = \tan^{-1}(k_j/v_j)$. In our experiment, the plane-parallel plate with thickness $d_i$ generate a lateral displacement to ray $Q_jR_j$. The displacement adds an error into the $k_j$. The displacement's maximum related to thickness $d_i$ is estimated using the formula (A6).

$$K_j - k_j \geq d_j \tan\theta_3 \quad (A6)$$

When $\theta_3 \approx 18.8$ and $d_i \approx 1$mm, $K_j - k_j \leq 340$μm. We select a large length, $v_j \geq 340$mm (i.e. $K_j$-$k_j$/$v_j$ $\leq 10^{-3}$) to reduce the error of ratio ($k_j/v_j$) related to the uncertainty of thickness $d_j$.

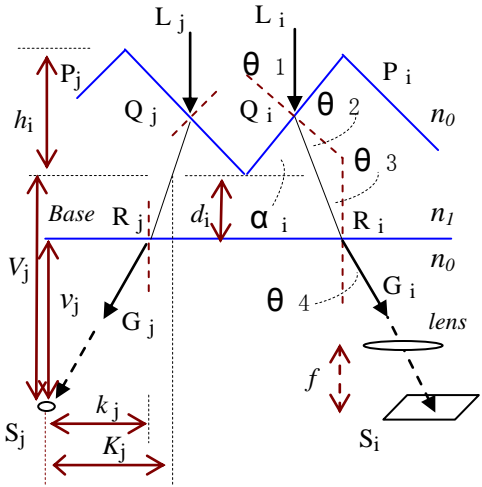

Fig.9 The refraction of light by a prism and the measuring of the dihedral angle $\alpha_i$

## *Acknowledgments*

This work was supported by Project of Innovation Practice and Research Foundation for Undergraduates, Chinese Hunan Province Education Department (Grant No.2010-5). The authors thank the Musen Technologies CO., LTD (Shenzhen, China) for experimental assistance.